\documentclass[sigconf]{acmart}

\AtBeginDocument{%
  }

\setcopyright{acmlicensed}
\copyrightyear{2026}
\acmYear{2026}
\acmDOI{XXXXXXX.XXXXXXX}

\acmConference[EASE 2026]{Proceedings of The 30th International Conference on Evaluation and Assessment in Software Engineering}{Tue 9 - Fri 12 June 2026}{Glasgow, United Kingdom}

\usepackage{amssymb}
\usepackage{graphicx}
\usepackage{subcaption}
\usepackage{booktabs}
\usepackage{multirow}
\usepackage[switch]{lineno}
\usepackage{xspace}
\usepackage{pifont}
\usepackage[export]{adjustbox}
\usepackage{amsmath}
\usepackage{svg}
\usepackage{tikz}
\usepackage{xcolor}
\usepackage{float}
\usepackage{mathtools}
\usepackage{pgfplots}
\usepackage{pgfplotstable}
\usepgfplotslibrary{groupplots}
\usepackage{float}
\usepackage{placeins}
\pgfplotsset{compat=1.18}
\usepackage[utf8]{inputenc}
\DeclareUnicodeCharacter{2264}{\ensuremath{\leq}}
\DeclareUnicodeCharacter{2265}{\ensuremath{\geq}}
\usetikzlibrary{calc}
\usetikzlibrary{matrix}
\usepgfplotslibrary{statistics}
\usetikzlibrary{arrows.meta,positioning,shapes.geometric}
\tikzset{
  boxframe/.style={x=1cm, y=0.06cm, line cap=round, line join=round},
  boxfill/.style={fill=orange!70!yellow, draw=black},
  gridline/.style={gray!60, dashed, line width=0.3pt},
  axisline/.style={black, line width=0.5pt}
}

\tikzset{boxfillA/.style={fill=orange!55, draw=orange!85!black}}
\tikzset{boxfillB/.style={fill=blue!55,   draw=blue!85!black}}
\tikzset{boxfillC/.style={fill=green!55,  draw=green!85!black}}
\tikzset{boxfillD/.style={fill=red!55,    draw=red!85!black}}

\tikzset{mypentagon/.pic={
    \draw[fill=gray!30,draw=black] (90:0.2cm)
    \foreach \x in {162,234,306,378} { -- (\x:0.2cm)} -- cycle;
}}

\ExplSyntaxOn
\NewDocumentCommand{\drawgrid}{m}{
  \group_begin:
  \clist_set:Nn \l__grid_sel_clist { #1 } 
  \begin{tikzpicture}[scale=0.8]
    \foreach \i in {0,...,3} {
      \foreach \j in {0,...,3} {
        \pgfmathtruncatemacro{\idx}{\i*4+\j}
        \tl_set:Nx \l__grid_idx_tl { \int_to_arabic:n{\idx} }
        \clist_if_in:NVTF \l__grid_sel_clist \l__grid_idx_tl
          { \filldraw[fill=gray!60, draw=black] (\j,-\i) rectangle ++(1,-1); }
          { \draw (\j,-\i) rectangle ++(1,-1); }
        \node at (\j+0.5,-\i-0.5) {\small \idx};
      }
    }
  \end{tikzpicture}
  \group_end:
}
\ExplSyntaxOff

\usepackage{ifthen}
\usepackage[normalem]{ulem} 
\usepackage{xcolor,colortbl}
\usepackage{hyperref}
\usepackage{subcaption}

\newboolean{showedits}
\setboolean{showedits}{true} 
\ifthenelse{\boolean{showedits}}
{
	\newcommand{\del}[1]{\textcolor{red}{\sout{#1}}} 
	\newcommand{\nbe}[3]{
		{\colorbox{#3}{\bfseries\sffamily\scriptsize\textcolor{white}{#1}}}
		{\textcolor{#3}{\sf\small$\blacktriangleright$\textit{#2}$\blacktriangleleft$}}}
}{
	\newcommand{\del}[1]{} 
	
	\newcommand{\nbe}[3]{}
}

\newboolean{showcomments}
\setboolean{showcomments}{true} 
\newcommand{\id}[1]{$-$Id: scgPaper.tex 32478 2010-04-29 09:11:32Z oscar $-$}

\ifthenelse{\boolean{showcomments}}
 {
 	\newcommand{\nbc}[3]{
 		{\colorbox{#3}{\bfseries\sffamily\scriptsize\textcolor{white}{#1}}}
		{\textcolor{#3}{\sf\small$\blacktriangleright$\textit{#2}$\blacktriangleleft$}}}
	
 }{
 	\newcommand{\nbc}[3]{}
 	
 }

\usepackage[most]{tcolorbox}
\ifthenelse{\boolean{showedits}}
{
  \newtcolorbox{inserted}{%
       title=Inserted text:,
       colframe=blue,colback=blue!5!white,
       breakable,
       leftrule=0mm, 
       bottomrule=0mm,
       rightrule=0mm,
       toprule=0mm,
       arc=0mm, outer arc=0mm,
       oversize
  }
  \newtcolorbox{deleted}{%
       title=Deleted text:,
       colframe=red,colback=red!5!white,
       breakable,
       leftrule=0mm, 
       bottomrule=0mm,
       rightrule=0mm,
       toprule=0mm,
       arc=0mm, outer arc=0mm,
       oversize
  }
  \newtcolorbox{refactored}{%
       title=Rewritten text:,
       colframe=blue,colback=red!5!white,
       breakable,
       leftrule=0mm, 
       bottomrule=0mm,
       rightrule=0mm,
       toprule=0mm,
       arc=0mm, outer arc=0mm,
       oversize
  }
}{

}
\newboolean{isblinded}
\setboolean{isblinded}{true}
\ifthenelse{\boolean{isblinded}}
{\newcommand\blind[1]{BLINDED\xspace}}
{\newcommand\blind[1]{#1\xspace}}

\newcommand{\commented}[1]{}

\newcommand{\eg}{\emph{e.g.,}\xspace}
\newcommand{\ie}{\emph{i.e.,}\xspace}

\definecolor{source}{gray}{0.9}

\begin{document}

\title{Pick and Sort for Graphical Authentication}

\author{Argianto Rahartomo}
\affiliation{
  \institution{Technische Universität Clausthal}
  \city{Goslar}
  \country{Germany}
}
\email{argianto.rahartomo@tu-clausthal.de}
\orcid{0000-0002-9592-0023}

\author{AmirHossein Jamshidipoor}
\affiliation{
  \institution{Islamic Azad University}
  \city{Tehran}
  \country{Iran}
}
\email{amirhossein.jamshidipoor@iau.ir}
\orcid{https://orcid.org/0009-0006-1302-9744}

\author{Mohammad Ghafari}
\affiliation{
  \institution{Tehran Institute for Advanced Studies (TEIAS), Khatam University}
  \city{Tehran}
  \country{Iran}
}
\email{m.ghafari@teias.institute}
\orcid{0000-0002-1986-9668}

\renewcommand{\shortauthors}{Rahartomo et al.}

\begin{abstract}
We propose a graphical authentication scheme that follows a simple  \emph{Pick and Sort} design in which users choose visual elements and arrange them within a grid. 
Both the number of selected elements and the grid size are configurable, and the visual elements can be customized for specific user groups, such as children.
A preliminary study with a prototype implementation indicated that the scheme is easy to learn and flexible to deploy. 
Although login times are longer than those of conventional authentication methods, the additional interaction may be acceptable in scenarios that are not time-critical, such as infrequent-access use cases or as a secondary authentication mechanism.
\end{abstract}




\maketitle

\section{Introduction}
\label{sec:introduction}
Authentication is a critical mechanism for protecting access to sensitive online services such as banking, e-commerce, education, telemedicine, and remote work systems.
Effective authentication mechanisms must balance security strength with usability, a challenge that continues to persist even with emerging technologies such as the metaverse~\cite{RAHARTOMO2025104602}.

Despite decades of technological progress, most authentication systems still rely on knowledge-based secrets such as passwords or PINs because they are inexpensive and straightforward to implement.
However, text-based passwords are often difficult to enter accurately, easy to forget, and easy to guess, which undermines security. 

Passkeys (WebAuthn/FIDO2) reduce the need to memorize passwords. However, they are bound to a specific website origin and are typically tied to a particular device or 
ecosystem. When users return after an extended period, they may have replaced their phone, lost a device, or removed an authenticator application. In such cases, access to the original passkey may no longer be available, requiring users to rely on fallback or account-recovery mechanisms.

Biometric authentication has also become a common local unlock mechanism. During authentication, the device verifies the biometric input locally and, upon success, uses the unlocked private key to generate a digital signature, which is then sent to the server. The server validates this signature using the corresponding public key, thus confirming the authentication attempt.
However, biometrics cannot operate as a standalone web authentication mechanism, as they depend on coordinated support from the device, operating system, browser, and service. Moreover, if the device or its associated credential is lost or replaced, users must again rely on fallback or account recovery procedures.

We propose a graphical authentication scheme called ``\emph{Pick and Sort}'' that integrates both recognition-based and recall-based components.
Users select a set of visual elements and arrange them within a grid, with each login requiring only a few simple clicks or touches.
%
Recognition helps users identify the correct elements without remembering specific visual details, which minimizes memory effort.
Recall is involved only in the ordering step, which adds structure to the secret and increases the number of unique arrangements, enhancing security.
The scheme is customizable in terms of the set of elements and the grid size, making it adaptable to different user groups and needs. Its visual and spatial layout provides an additional layer of security, as an attacker must know both the chosen elements and their exact positions on the grid to succeed. This method is platform-independent, relying only on basic interface components widely supported by most devices.

We developed a proof-of-concept implementation of this scheme and conducted a preliminary user study.
We measured the login success rate, the time spent on registration and authentication, and the time required for each step of element selection.
We also collected the most frequently chosen elements, the elements that tend to co-occur, and how users spatially arrange elements within the grid.
The results provide early, exploratory insights into how participants interact with the scheme.
%
On average, they complete the registration process in under a minute. We observe distinct differences in the ways users select and arrange their elements on the grid, with minimal overlap in their choices. In addition, we find that login success improves with practice, even after a gap of weeks between sessions, while the overall interaction time remains consistent.

Although the average login time is longer than that of conventional authentication methods, the scheme remains a viable option for scenarios that do not require time-critical access and where logins occur infrequently. We do not aim to replace existing approaches such as passkeys, biometrics, or multi-factor authentication, but rather to provide a complementary method supported by preliminary evidence for infrequent-access use cases. 
This is particularly relevant in browser-based settings where advanced authentication options may not be universally available, fully deployed, or consistently supported, and where account-recovery mechanisms vary widely across services.
The scheme may also be suitable for specific user groups, such as children and older adults, who may perceive the interaction as game-like while benefiting from improved protection against weak or reused passwords. However, 
real-world studies
are needed to 
assess
long-term usability and validate these potential benefits.
%

The remainder of this paper is organized as follows.
Section~\ref{sec:relatedwork} presents background information and related studies.
Section~\ref{sec:designing} describes the design of our proposed graphical authentication scheme.
Section~\ref{sec:evaluation} outlines the experimental setup and presents the results.
Section~\ref{sec:ttv} discusses potential threats to validity, and finally, Section~\ref{sec:conclusion} concludes the paper.

\section{Background and Related Work}
\label{sec:relatedwork}

We give an overview of authentication methods and then present graphical authentication systems available in the literature.

\subsection{Authentication}
\label{subsec:authingeneral}

%

Authentication is a fundamental security service that verifies the identity of an entity requesting access to protected resources. The entity may be a person, device, computer, or system, and authentication ensures that only legitimate users can access sensitive data or perform critical operations. This verification is typically based on one or more authentication factors: something the user knows, something the user possesses, or something the user is.

\emph{Something you know/they know (or Knowledge-based):} This includes passwords, passphrases, and personal identification numbers (PINs).
It requires users to remember and enter a secret, such as a password or the answer to a security question.
The advantage of this method is simple, low cost, and universal support across various digital platforms.
Users are generally familiar with how to use it, and it does not require any special hardware.
However, it has some disadvantages, particularly in terms of memorability, especially for complex, long, and unique passwords.
In addition, it is highly susceptible to threats such as guessing, phishing, brute-force attacks, and database breaches.
Many users tend to reuse weak passwords, which is not considered a good security practice;

\emph{Something you have/they have (or Possession-based):} This method relies on something the user physically possesses for authentication, such as a USB token, a mobile phone that generates one-time passwords (OTPs), or a smartcard.
It provides strong resistance against phishing and replay attacks and typically does not require users to memorize anything.
However, users must carry an additional item, which can be lost, stolen, or damaged.
Furthermore, it incurs additional deployment costs and may present logistical challenges, especially in large-scale systems or diverse user environments.

\emph{Something you are/they are (or Biometric-based):} Inherence-based authentication utilizes biometric characteristics unique to each user.
It offers convenience, as users do not need to remember or carry anything.
The authentication process is generally fast and intuitive.
Despite these usability advantages, this method can encounter challenges like false positives and false negatives, particularly in noisy or variable conditions.
Additionally, biometric data cannot be changed like a password if compromised, and not all devices may have the necessary hardware for this type of authentication;


\subsection{Graphical Authentication}
\label{subsec:graphicalauth}

Graphical authentication schemes are commonly classified according to how users interact with visual elements during the login process, with three main types frequently identified in the literature~\cite{biddle2012graphical}.

%
%
%
%
%
%

\emph{Recognition-based} schemes ask users to recognize and select one or more images from a fixed pool.
%
During registration, users choose a small set of familiar or meaningful images, such as faces, icons, or objects.
During login, they must find and select the same images again among several decoy images.
%
These schemes use the human ability to recognize visual patterns.
They often show good memorability and require little effort because users only need to spot known images instead of reproducing a pattern from memory.
%
However, recognition-based schemes usually offer limited entropy.
Many users tend to choose similar or ``obvious'' images, and the fixed pool size restricts the number of distinct secrets.
Attackers can exploit this bias and try the most popular images first, making targeted guessing attacks easier, especially when the same global image pool appears across services.
%
Passfaces~\cite{alameen2015} is an example of a recognition-based graphical password that uses human faces as the main image pool.
%

\emph{Recall-based} schemes require users to reproduce a visual password from memory without any visual cues during the login process.
%
In these schemes, users create a secret by drawing a pattern or performing a sequence of actions on a predefined interface. A common example is the Android Pattern Unlock (APU), where users connect dots on a 3$\times$3 grid to form a pattern. During login, they must replicate the same pattern on the same grid.
%
Recall-based schemes offer a large password space since many different patterns are possible. As a result, they theoretically provide high security when users select patterns randomly.
%
However, usability issues often arise with these schemes. Users may forget the exact pattern or confuse similar patterns, especially after a period of infrequent use. Small variations in drawing can also lead to errors, such as when users lift their finger too early or connect nodes in the wrong order.
%
Bu-Dash~\cite{panagiotis2023} is a variant of APU that combines shape-based input with randomization. It employs a 3$\times$3 grid, but each node displays geometric shapes (such as \ding{109}, $\triangle$, $\times$, --, or $\square$),  which change with each interaction. To create a password, users connect nodes in a specific sequence of shapes (e.g., $\triangle$--$\times$--) instead of using fixed node positions. This design retains the familiar grid layout while enhancing resistance to observation attacks. However, it focuses on single-user, single-device scenarios and does not accommodate use across multiple devices or services. 
%
%

\emph{Cued recall-based} schemes assist users in remembering their passwords by providing visual cues during the authentication process.
%
Unlike traditional recall, where users depend solely on their memory, cued recall schemes incorporate background images or contextual elements to aid users in reconstructing their passwords. For example, in PassPoints~\cite{dirik2007}, users select a sequence of points on a background image during registration and must click the same points when logging in. The image acts as a memory aid, helping users locate the correct points again.
%
However, cued recall-based schemes encounter specific security and usability challenges. Users often select visually prominent ``hotspots'', such as notable objects or corners, which reduces the effective password space and makes it easier for attackers to guess passwords through targeted guessing. In addition, these schemes usually require precise clicking; minor deviations from the original points can result in login failures due to strict tolerance thresholds.

Therefore, existing graphical authentication schemes may be technology-dependent, limited to single-device scenarios, prone to predictable visual hotspots, or reliant on complex multi-step gestures.

\section{The Pick and Sort Scheme}
\label{sec:designing}

%

\begin{figure*}[h!]
\centering
\resizebox{\textwidth}{!}{%
\begin{tikzpicture}[scale=1]
  
  \draw (0,0) grid ++(3,2);
  \node[below, align=center] at (1.5,0) {\small{Sets of elements}};

  \node at (0.5,1.5){\includegraphics[width=0.5cm]{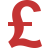}};
  \node at (0.5,0.5){\includegraphics[width=0.5cm]{img/currency-2.png}};

  \node at (1.5,1.5){\includegraphics[width=0.5cm]{img/human-1.png}};
  \node at (1.5,0.5){\includegraphics[width=0.5cm]{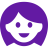}};

  \node at (2.5,1.5) {\includegraphics[width=0.5cm]{img/construction-1.png}};
  \node at (2.5,0.5){\includegraphics[width=0.5cm]{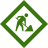}};

  \draw (4,0) grid ++(3,3);
  \node[below, align=center] at (5.5,0) {\small{First choice}};
  \fill[gray!50] (4,2) rectangle ++(1,1);
  \node at (4.5,2.5){\includegraphics[width=0.5cm]{img/human-1.png}};

  \draw (8,0) grid ++(3,3);
  \node[below, align=center] at (9.5,0) {\small{Second choice}};
  \node at (8.5,2.5){\includegraphics[width=0.5cm]{img/human-1.png}};
  \fill[gray!50] (9,1) rectangle ++(1,1);
  \node at (9.5,1.5){\includegraphics[width=0.5cm]{img/currency-2.png}};

  \draw (12,0) grid ++(3,3);
  \node[below, align=center] at (13.5,0) {\small{Third choice}};
  \node at (12.5,2.5){\includegraphics[width=0.5cm]{img/human-1.png}};
  \node at (13.5,1.5){\includegraphics[width=0.5cm]{img/currency-2.png}};
  \fill[gray!50] (14,0) rectangle ++(1,1);
  \node at (14.5,0.5){\includegraphics[width=0.5cm]{img/construction-1.png}};

\end{tikzpicture}%
}
\caption{Overview of the workflow for the ``Pick and Sort'' authentication scheme.}
\label{fig:authworkflow}
\end{figure*}

We introduce \emph{Pick and Sort}, a graphical authentication scheme that uses a grid-based interface together with a set of visual elements for user authentication.
Figure~\ref{fig:authworkflow} illustrates the overall scheme and its workflow.
The specific parameters shown (three element sets, two elements per set, and a 3×3 grid) are illustrative only.
First, the user selects the human icon \includegraphics[width=0.35cm]{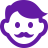}
and places it in the top-left cell of the grid. 
Next, the user selects the currency icon \includegraphics[width=0.35cm]{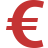}
and places it in the center cell. 
Finally, the user chooses the truck icon \includegraphics[width=0.35cm]{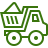}
and places it in the bottom-right cell. 
These three elements, arranged diagonally within the grid, together form the user's authentication secret.

The four main characteristics of this scheme are described below, based on previous works~\cite{bonneau2012, noahdas2025, katsini2025}.

\begin{itemize}

    \item \emph{Effectiveness} refers to how well an authentication scheme can correctly verify legitimate users and reject impostors. 
    The recognition component helps legitimate users correctly identify their chosen elements~\cite{biddle2012graphical,meng2025}, while the recall component supports remembering by requiring them to place these elements on the grid in the same way each time.

    \item \emph{Usability} refers to how easy, efficient, and satisfying the authentication scheme is for users.
    %
    %
    Through this scheme, users follow a straightforward process of selecting images and arranging them visually, which is more engaging than typing PINs or passwords. Each login attempt involves just a few simple clicks or touches, and users can easily reselect elements or adjust their arrangement, making the process faster and more efficient. This scheme treats the secret as the final configuration of the selected elements, and it does not use the selection or placement order as part of the secret. We keep the scheme order-independent because order increases cognitive burden. In addition, the elements and the grid size can be customized for different user groups, such as based on region, age, or professions, which enhances ease of use and overall satisfaction.

    \item \emph{Security} refers to the ability to withstand various types of attack.
    Due to the visual and spatial nature of the scheme, even if an attacker observes or guesses the chosen elements, they still need to know the exact position of each element within the grid. Implementing policies such as a minimum number of element selections, offering multiple element sets, and increasing the grid size exponentially expands the password search space, thereby reducing the risk of brute-force or guessing attacks.
    The actual entropy depends on the choices users make when selecting and arranging elements.
    Nonetheless, the upper limit entropy of this scheme (\ie theoretical entropy) would be calculated as shown below.

    We suppose $n$ is the grid size and the user selects $k$ elements that should be placed within the grid. These elements belong to $m$ disjoint sets. Hence, the total number of elements would be:
    \[
    S = \sum_{i=1}^{m} s_i
    \]

    There are $\binom{n}{k}$ ways to choose the occupied cells, and each occupied cell can independently contain one element from the total pool $S$, allowing repetition, under the condition that each subset $S_i$ must appear at least once. Therefore, based on the inclusion-exclusion principle~\cite{berman1972, kahn1996}, the number of valid labelings are:
    
    \begin{equation}
    M(k) =
    \sum_{j=0}^{m} (-1)^j
    \sum_{1 \le i_1 < \cdots < i_j \le m}
    \left(S - \sum_{t=1}^{j} s_{i_t}\right)^k
    \end{equation}

The number of valid arrangements for $k$ elements are:

    \begin{align}
    N_k &= \binom{n}{k} \, M(k)
    \end{align}

Finally, this would result in the following total password space:

    \begin{align}
    N &= 
    \sum_{k=k_{\min}}^{k_{\max}} \binom{n}{k}\, M(k)
    \label{eq:entropy_general}
    \end{align}

    The entropy expression in our formulation illustrates the theoretical password space that the scheme can support when users are allowed to select any value of $k$ within the permitted range.
    
    \item \emph{Deployability} refers to the ease of implementation in a technology-agnostic manner.
    %
    Our scheme can be implemented on multiple platforms, including web browsers, mobile devices, and desktop applications.
    It utilizes a smaller grid and simple tap actions to enable users to easily select elements. We also implement adaptive layouts that adjust cell sizes and spacing to accommodate different screen sizes.
    The scheme requires minimal computational resources and no specialized hardware or software. 
    Therefore, this lightweight and technology-independent design is highly deployable and easily adaptable to various environments.
    
\end{itemize}


In summary, ``Pick and Sort'' 
has the potential to offer a simple, usable, and robust authentication scheme that is easy to deploy and adaptable across diverse demographics and contexts.


\section{Evaluation}
\label{sec:evaluation}

We describe our study design and prototype development and then present the study results 
as early-stage evidence that characterizes the interaction behavior and trade-offs of this scheme.

\subsection{Study Design}

\emph{The prototype.}
We developed the prototype within the scope of a website authentication system.
It uses HTML, CSS, and JavaScript for the front end, Django framework for the back end, and SQLite for data management.

\begin{figure}[h]
    \centering
    \resizebox{\columnwidth}{!}{%
        \begin{tabular}{ccc}
            \includegraphics[width=0.3\textwidth]{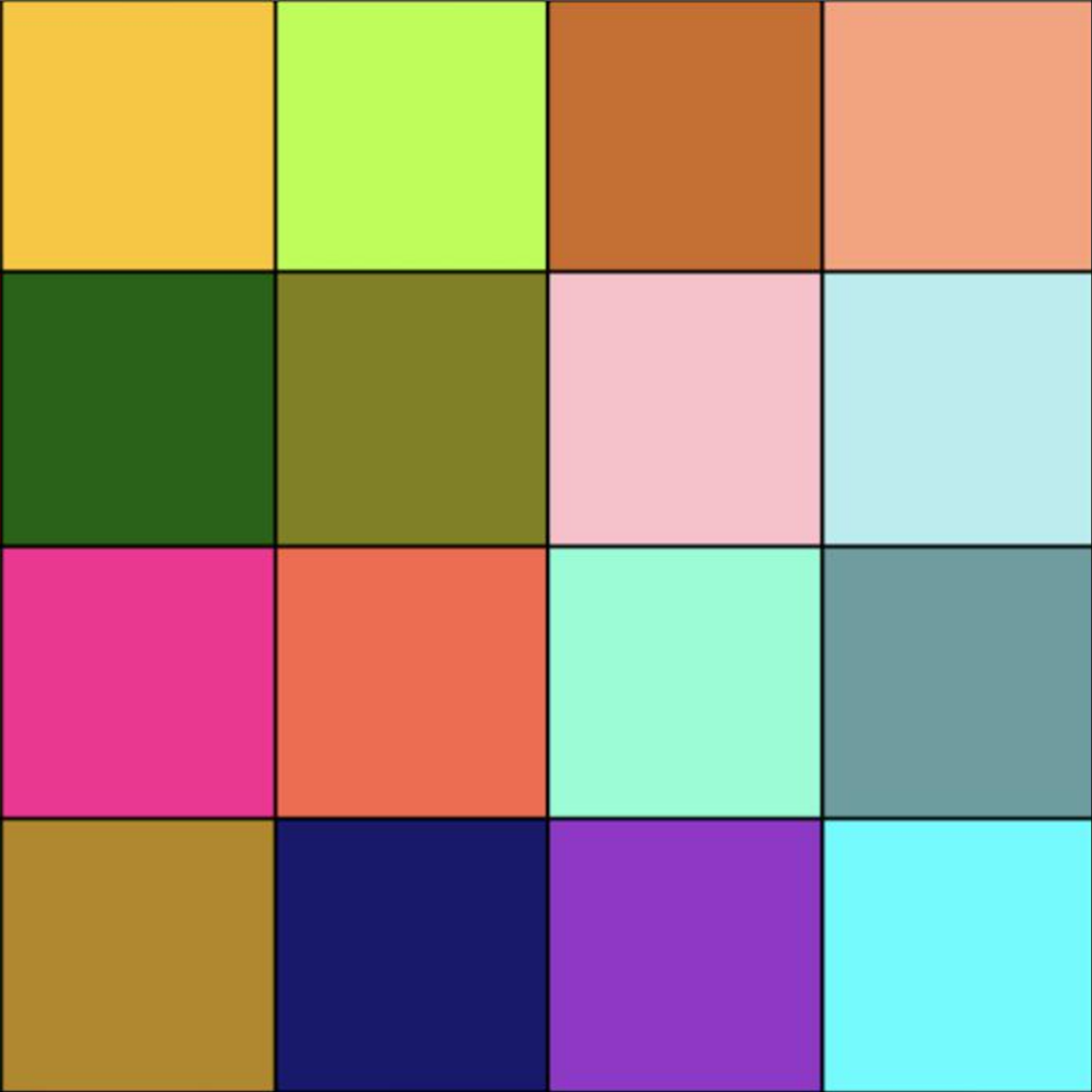} &
            \includegraphics[width=0.3\textwidth]{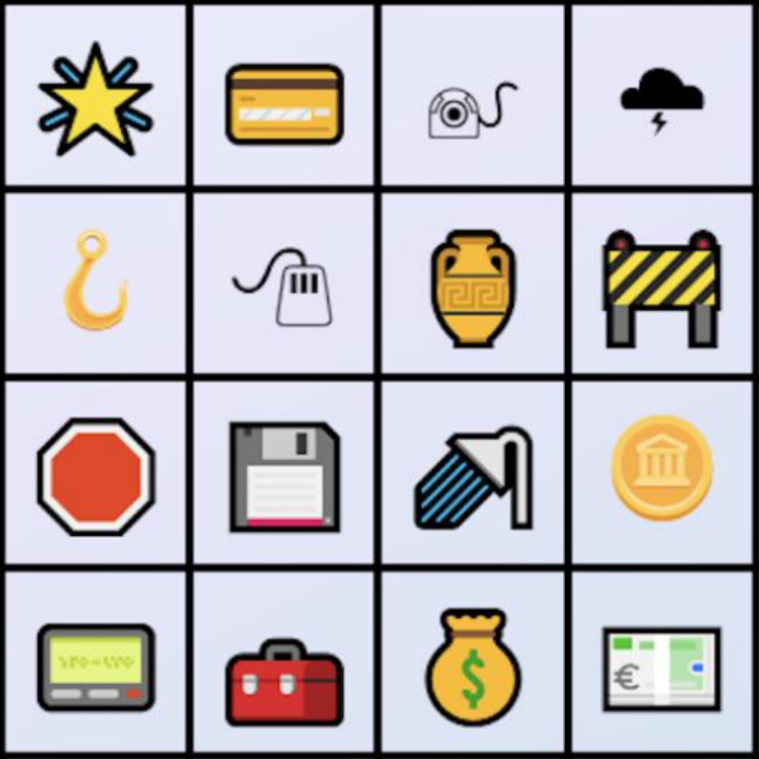} &
            \includegraphics[width=0.3\textwidth]{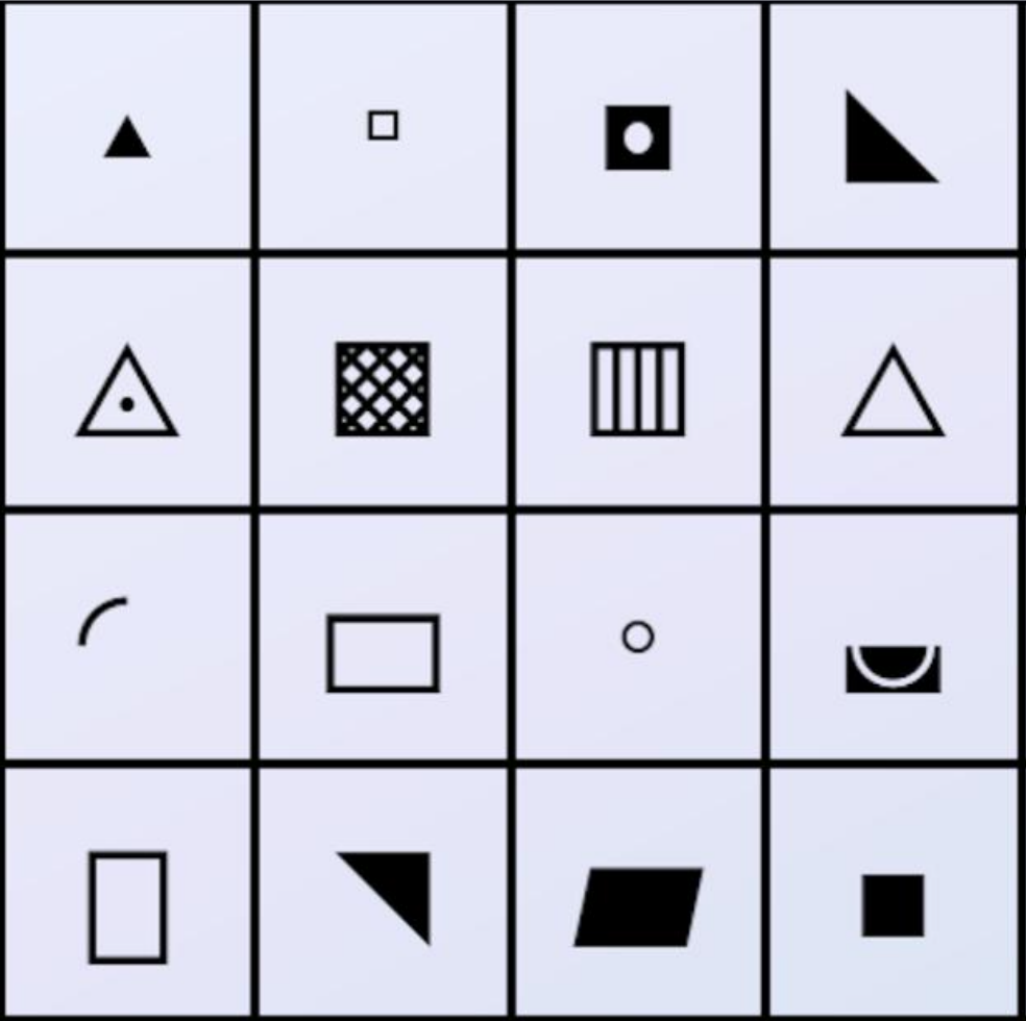}
        \end{tabular}%
    }
    \caption{Excerpt of three element sets in the prototype.}
    \label{fig:palettes}
\end{figure}

We instantiated the system with three sets of elements, namely colors (40), icons (90), and shapes (50), where the numbers in parentheses indicate the size of each set. 
We manually curated all three sets of elements for this prototype and did not depend on any external datasets. For the color set, we recognize that some users may struggle to distinguish colors as the pool grows large (\eg 20 or more). Currently, our configuration does not include a color-blind–friendly palette, but the setting can be changed to use a smaller and more discriminable palette if desired.
We randomize the positions of all elements each time a user logs in to reduce the risk of guessing attacks and to mitigate position-based shortcuts that could emerge when a fixed layout is repeated. The prototype uses within-set randomization: it displays the three sets and shuffles only the positions of elements within each set. Each element remains in its original set and does not randomize elements between sets, which preserves set semantics while still varying the on-screen layout between attempts.
Figure~\ref{fig:palettes} shows an excerpt of each set. 

We report effectiveness in terms of legitimate-user performance (\eg login success) and do not claim empirically measured impostor rejection. While for usability, we focus on behavioral usability indicators (\eg task completion, errors, and time) and do not claim subjective satisfaction, liking, or recommendation because we did not collect post-study questionnaire data. For security, we do not present a realistic attack evaluation (\eg observation studies or adversarial guessing) and, therefore, do not claim measured resistance against attackers. Instead, we discuss security in a scoped manner by characterizing the secret representation and reporting sample-based indicators of choice diversity (\eg secret count) to inform predictability and design trade-offs; these indicators are not security guaranties. 
Finally, for deployability, our prototype demonstrates feasibility using basic user interface (UI) components, and we do not claim validated performance across diverse devices or environments. 

Our study evaluates a single prototype implementation. All participants use the same prototype, and there is no assignment of participants to different prototypes. A facilitator provides participants with brief verbal instructions on when to start, and the application displays an on-screen “Login” label at the entry point.
%


We enforce the policy that users select minimum one element from each set, and sort them within a 4×4 grid.
The maximum number of selections is 16, which is equal to the grid size.
%
%
The upper bound entropy of the configuration results is nearly 120 bits (equation~\ref{eq:entropy_general} references the precise calculation), exceeding Bonneau’s recommended threshold~\cite{bonneau2012}. 
However, this value does not reflect the actual secrets generated in our pilot study, as the true entropy depends on the choices users make when selecting and arranging elements.

The system provides feedback on whether a login attempt was successful.
For unsuccessful logins, users could try again without losing any previously entered data, allowing them to review and adjust their earlier selections without redoing the entire process.

\emph{The study.}
We invited 82 participants to this study through personal connections and the university community.
We explained the purpose of this experiment to the participants and provided them with a tutorial that explains the authentication approach.
We explicitly informed them of the data that we collect such as login attempts, user choices and patterns, login durations, and we gathered their consent prior to the study.

Participation is completely voluntary, with no payment or compensation provided, and participants could withdraw at any time. We obtained informed consent from all participants before they took part in the study. For participants under 18 years old, we also obtained consent from a parent or legal guardian, as well as the child’s assent when applicable. To protect privacy, we collected only non-identifying data required for the analysis, including age and interaction logs. We did not collect additional demographic information such as gender, education level, or field of study.

Each participant had to register once and ideally complete three login attempts, scheduled for one day after registration (\ie first login stage), ten days after registration (\ie second login stage), and twenty-eight days after registration (\ie third login stage). We consider longer delays in our future studies.

There were 59 participants who proceeded to the first login stage. The majority, \ie 36 participants, were aged between 20 and 30 years; 13 participants were older than 30, and the remaining 10 participants were under 20. The sample consists of both adults and minors. For participants under 18 years old, we obtain consent from a parent or legal guardian, as well as the child’s assent when applicable. Within the group of participants under 20 years old, there are 10 minors (those under 18), including 5 participants who are younger than 10 years old. 


\subsection{Results}
\label{sec:resultanddiscussion}

We present performance measures, including registration and login times and success rates, along with observations on element selection and, finally, the arrangements within the grid.

We focus exclusively on the behavior of genuine users in our controlled sessions and do not assess impostor rejection or draw conclusions about false acceptance or rejection rates. Evaluating impostor behavior is reserved for future studies with a different design.

\subsubsection{Performance Measures}
Out of the 59 participants, nine dropped out after their first login, and an additional seven discontinued after the second login. This left 43 participants whose login attempts were recorded across three stages.
%
Therefore, the retention rate (\ie the proportion of participants who complete all phases versus those who only enter the first phase) would be 72.88\%.

\begin{figure}
\centering
\begin{tikzpicture}
\begin{axis}[
  width=0.47\textwidth,  
  height=5.8cm,
  ymin=0, ymax=120,
  ymajorgrids,
  grid style={solid,gray!30},
  ylabel={Registration time},
  xmin=0.5, xmax=1.5,
  xtick=\empty,
  boxplot/draw direction=y,
  boxplot/box extend=0.10,
  every axis plot/.append style={solid},
  boxplot/every box/.style={draw=black, thick, solid},
  boxplot/every whisker/.style={draw=black, very thick, solid},
  boxplot/every median/.style={draw=black, very thick, solid},
  tick label style={font=\scriptsize},
  label style={font=\scriptsize},
  ytick={0,20,40,60,80,100,120},
]

\addplot+[
  fill=red!45,
  boxplot prepared={
    draw position=0.90,
    lower whisker=19,
    lower quartile=28,
    median=34,
    upper quartile=45,
    upper whisker=47
  }
] coordinates {};

\addplot+[
  fill=blue!45,
  boxplot prepared={
    draw position=1.05,
    lower whisker=9,
    lower quartile=28,
    median=51,
    upper quartile=66,
    upper whisker=101
  }
] coordinates {};

\addplot+[
  fill=green!45,
  boxplot prepared={
    draw position=1.20,
    lower whisker=24,
    lower quartile=28,
    median=55,
    upper quartile=89,
    upper whisker=115
  }
] coordinates {};

\end{axis}

\node[anchor=north] at ($(current bounding box.south)+(0,0cm)$) {%
  \begin{tikzpicture}[scale=0.9, baseline]
    \draw[fill=red!45,  draw=black] (0,0) rectangle +(0.35,0.15);
    \node[anchor=west, font=\scriptsize] at (0.50,0.08) {$<$20 yrs};
    \draw[fill=blue!45, draw=black] (2.0,0) rectangle +(0.35,0.15);
    \node[anchor=west, font=\scriptsize] at (2.50,0.08) {20--30 yrs};
    \draw[fill=green!45,draw=black] (4.0,0) rectangle +(0.35,0.15);
    \node[anchor=west, font=\scriptsize] at (4.50,0.08) {$>$30 yrs};
  \end{tikzpicture}
};

\end{tikzpicture}
\vspace{-0.9em}
\caption{Registration times by age groups.}
\label{fig:regtime-boxplots2}
\end{figure}
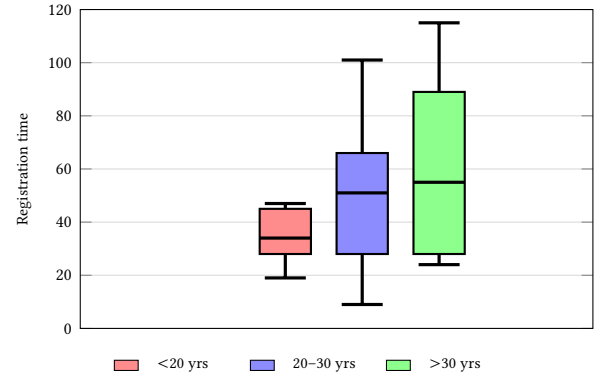

The average time spent for registration was about 50 seconds (standard dev 27.783 seconds).
Younger participants seem to be faster in registration than older participants (see Figure~\ref{fig:regtime-boxplots2}).
They were also faster in the initial login attempts, but the difference decreased by the third stage (see Table~\ref{tab:agegroups-stats}).
In other words, repeated login experience reduced the overall login time.
%
%
We include these larger pools to support diversity of element choice, but we acknowledge that increasing set size can also increase visual search and recognition demands during both registration and login. In particular, distinguishing many hues and scanning a large icon set may be challenging for some users and may contribute to longer entry times, suggesting that set size and palette design are important configuration choices.

\begin{table}
\centering
\caption{Authentication times by age groups.}
\label{tab:agegroups-stats}
\renewcommand{\arraystretch}{1.15}
\setlength{\tabcolsep}{5pt}
\begin{tabular}{lccccccc}
\toprule
\multirow{2}{*}{\textbf{Login Stage}} & 
\multicolumn{2}{c}{\textbf{$<$20 yrs}} &
\multicolumn{2}{c}{\textbf{20--30 yrs}} &
\multicolumn{2}{c}{\textbf{$>$30 yrs}}\\ 
\cmidrule(lr){2-3} \cmidrule(lr){4-5} \cmidrule(lr){6-7}
& \textbf{Mean} & \textbf{SD} & \textbf{Mean} & \textbf{SD} & \textbf{Mean} & \textbf{SD} \\
\midrule
Login-1  & 17.39  & 11.54  & 20.58 & 12.36 & 31.93 & 10.91 \\
Login-10 & 24.63  & 18.53  & 23.01  & 15.56 & 28.33 & 15.25 \\
Login-28  & 27.81  & 15.04  & 17.64  & 8.27  & 25.87 & 11.6 \\
\bottomrule
\end{tabular}
\end{table}

\begin{figure}[t]
\centering
\begin{tikzpicture}
\begin{axis}[
  width=0.47\textwidth,
  height=5.2cm,
  ymin=0, ymax=100,
  ymajorgrids,
  grid style={solid,gray!30},
  ylabel={Success rate (\%)},
  xmin=0.5, xmax=1.5,
  xtick=\empty,
  boxplot/draw direction=y,
  boxplot/box extend=0.10,
  every axis plot/.append style={solid},
  boxplot/every box/.style={draw=black, thick, solid},
  boxplot/every whisker/.style={draw=black, very thick, solid},
  boxplot/every median/.style={draw=black, very thick, solid},
  tick label style={font=\scriptsize},
  label style={font=\scriptsize},
  ytick={0,25,50,75,100},
]


\addplot+[
  fill=red!45,
  boxplot prepared={
    draw position=0.90,
    lower whisker=16.67,
    lower quartile=33.33,
    median=50,
    upper quartile=100,
    upper whisker=100
  }
] coordinates {};

\addplot+[
  fill=blue!45,
  boxplot prepared={
    draw position=1.05,
    lower whisker=50,
    lower quartile=66.67,
    median=66.67,
    upper quartile=83.335,
    upper whisker=100
  }
] coordinates {};

\addplot+[
  fill=green!45,
  boxplot prepared={
    draw position=1.20,
    lower whisker=14.29,
    lower quartile=58.57,
    median=80,
    upper quartile=100,
    upper whisker=100
  }
] coordinates {};

\end{axis}

\node[anchor=north] at ($(current bounding box.south)+(0,0cm)$) {%
  \begin{tikzpicture}[scale=0.9, baseline]
    \draw[fill=red!45,  draw=black] (0,0) rectangle +(0.35,0.15);
    \node[anchor=west, font=\scriptsize] at (0.50,0.08) {Login-1};

    \draw[fill=blue!45, draw=black] (2.0,0) rectangle +(0.35,0.15);
    \node[anchor=west, font=\scriptsize] at (2.50,0.08) {Login-10};

    \draw[fill=green!45,draw=black] (4.0,0) rectangle +(0.35,0.15);
    \node[anchor=west, font=\scriptsize] at (4.50,0.08) {Login-28};
  \end{tikzpicture}
};

\end{tikzpicture}
\vspace{-0.9em}
\caption{Login success rates.}
\label{fig:boxplots-success}
\end{figure}
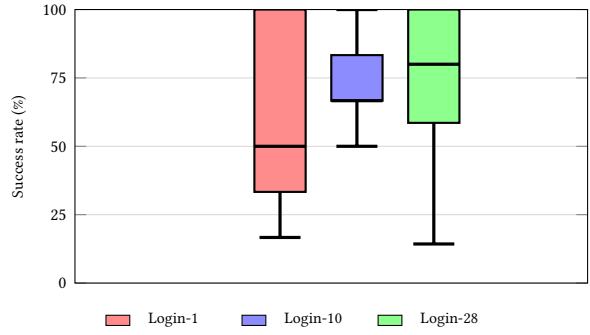

In total, we recorded 252 login attempts (mean 4.5, sd 2.23). The login activity for 21 participants was always successful.
We calculated the login success rate by dividing the number of successful attempts by the total number of attempts made by each participant (see Figure~\ref{fig:boxplots-success}).
The average success rate in the first stage was 58\% (sd 33.33\%), but it increased to 74\% (sd 18.897\%) in subsequent stages.
This indicates that participants achieved more successful logins in later stages, even when logging in 28 days after the registration phase.
The 1-, 10-, and 28-day gaps provide preliminary results, and we plan to conduct longer retention validation in our future studies.
Nonetheless, the authentication time for successful attempts across the three stages remained constant (mean 22 and sd 12 seconds).
This duration is longer than that of typical text-based authentication methods. 
We hypothesize that the randomization of element positions in each login attempt, implemented in our prototype, may have contributed to this. While we adopted this design to enhance security, it requires users to actively search for their selected elements rather than relying on fixed, memorized locations.

We conclude that the scheme becomes easier for users with practice, is reasonably memorable after a month, and supports successful long-term use. However, the initial onboarding experience could be improved to reduce failures in the first stage.

\subsubsection{Element Selection}

Most participants selected three to five elements (mean 3.96, sd 1.37), although about 20\% selected more \ie six to eight elements.
We noted that younger participants tend to select fewer elements than others. 
We observe that users who select more elements (indicated by a larger $k$) can take longer to log in (Spearman $\rho = 0.41$, $p < 0.001$, $N = 59$). 
In contrast, the login stage itself does not appear to affect the duration of the login process.

Table~\ref{tab:elementfreq} presents the \emph{most frequently selected element types}.
More than 50\% were colors, with black being the most commonly chosen color.
The proportions of selected icons and shapes were relatively balanced.
Among the color choices, black was selected the most frequently, appearing 38 times, followed by red (15 times), blue (13 times), Medium Slate Blue (9 times), and Dark Green (8 times). 

\begin{table}[htbp]
\centering
\scriptsize
\caption{Frequently selected elements.}
\begin{tabular}{llc}
\hline
\textbf{Element Type} & \textbf{Element} & \textbf{Frequency} \\
\hline
\multicolumn{3}{c}{\textit{Colors}} \\
\hline
Color & \colorbox[HTML]{000000}{\phantom{XX}} Black & 38 \\
 & \colorbox[HTML]{FF000F}{\phantom{XX}} Red & 15 \\
 & \colorbox[HTML]{0000FF}{\phantom{XX}} Blue & 13 \\
 & \colorbox[HTML]{7B68EE}{\phantom{XX}} MediumSlateBlue & 9 \\
 & \colorbox[HTML]{006400}{\phantom{XX}} DarkGreen & 8 \\
\hline
\multicolumn{3}{c}{\textit{Icons}} \\
\hline
Icon & \includegraphics[height=1em]{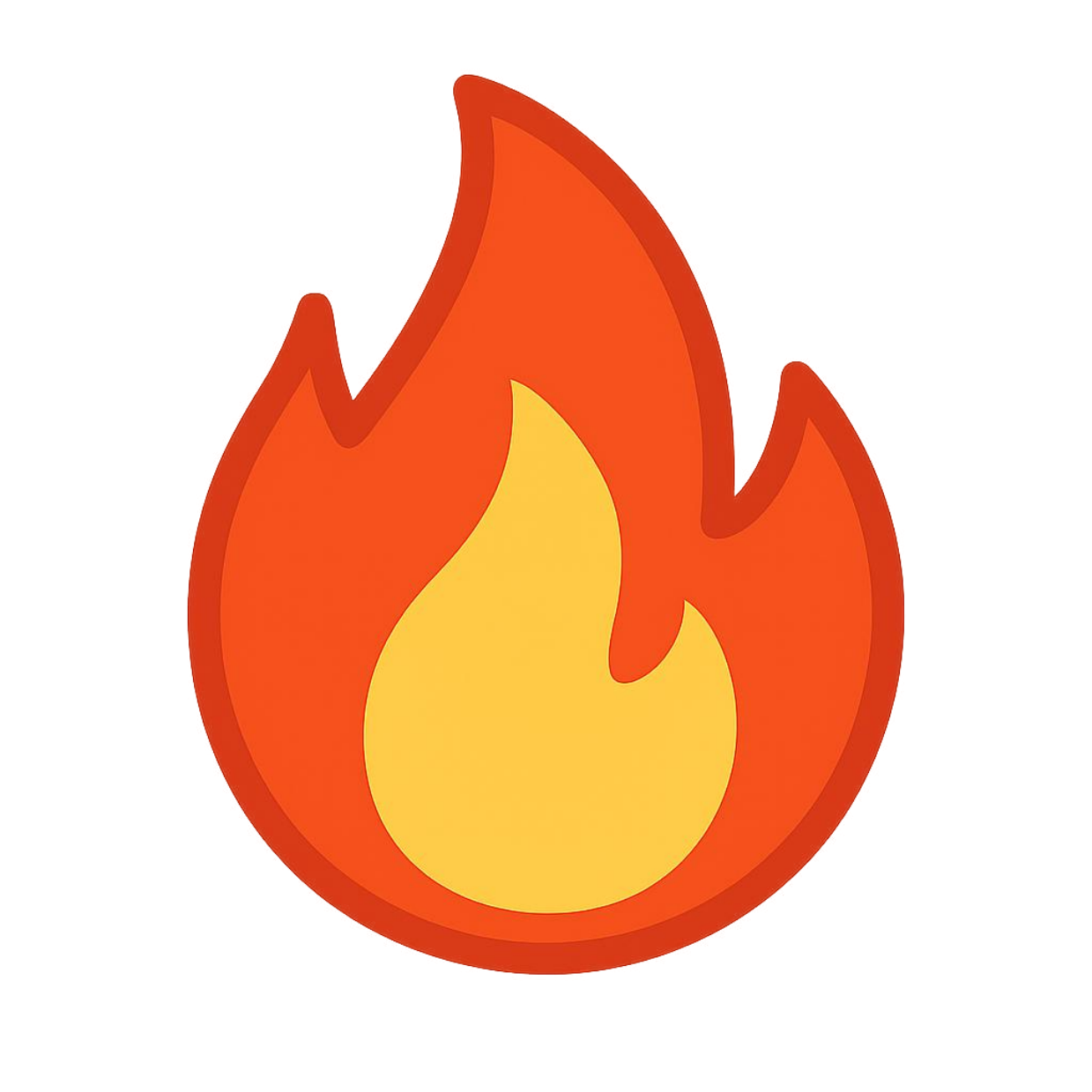} Fire & 10 \\
 & \includegraphics[height=1em]{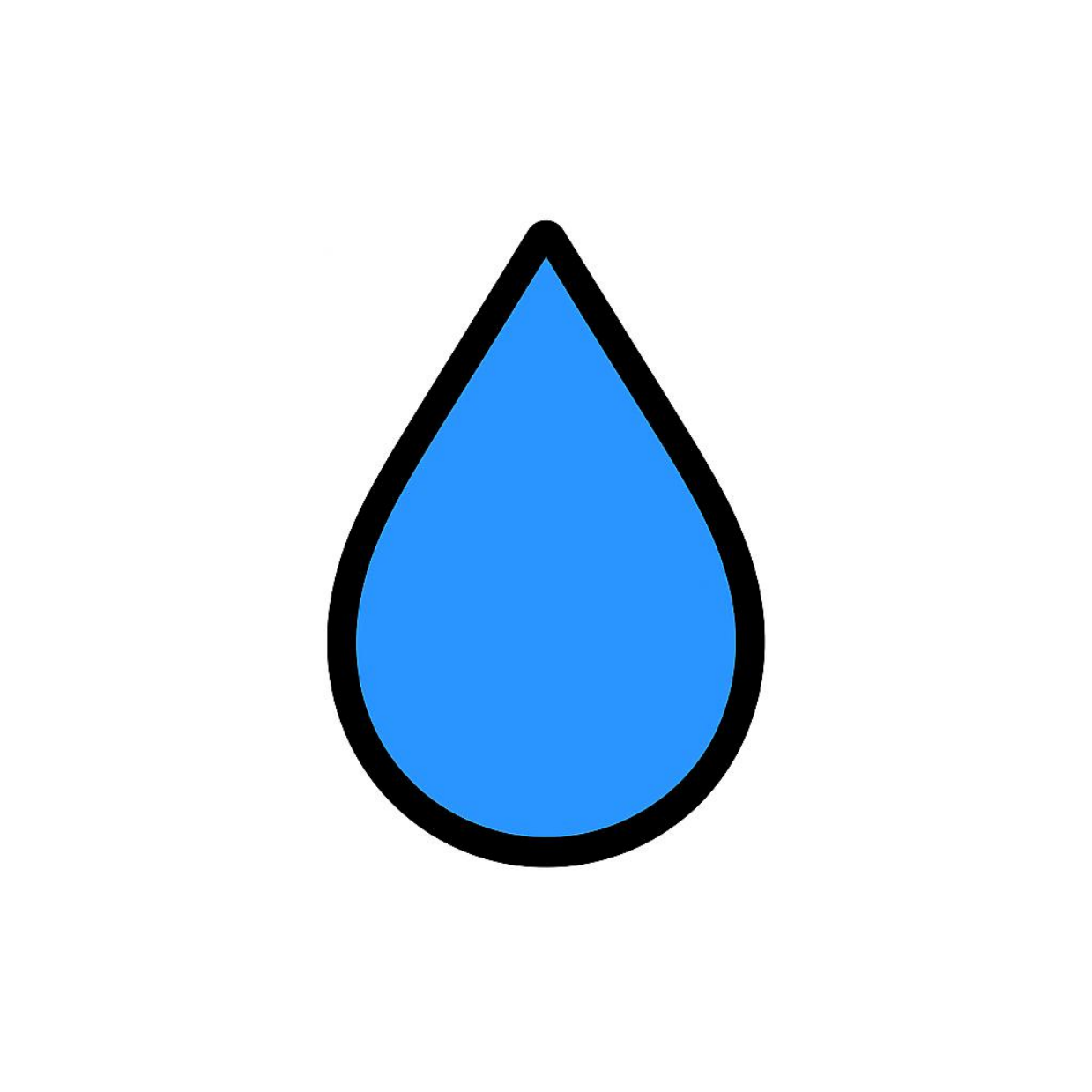} Water Droplet & 8 \\
 & \includegraphics[height=1em]{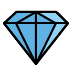} Gem Stone & 7 \\ 
 & \includegraphics[height=1em]{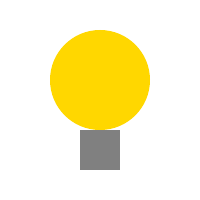}  Light Bulb & 6 \\ 
 & \includegraphics[height=1em]{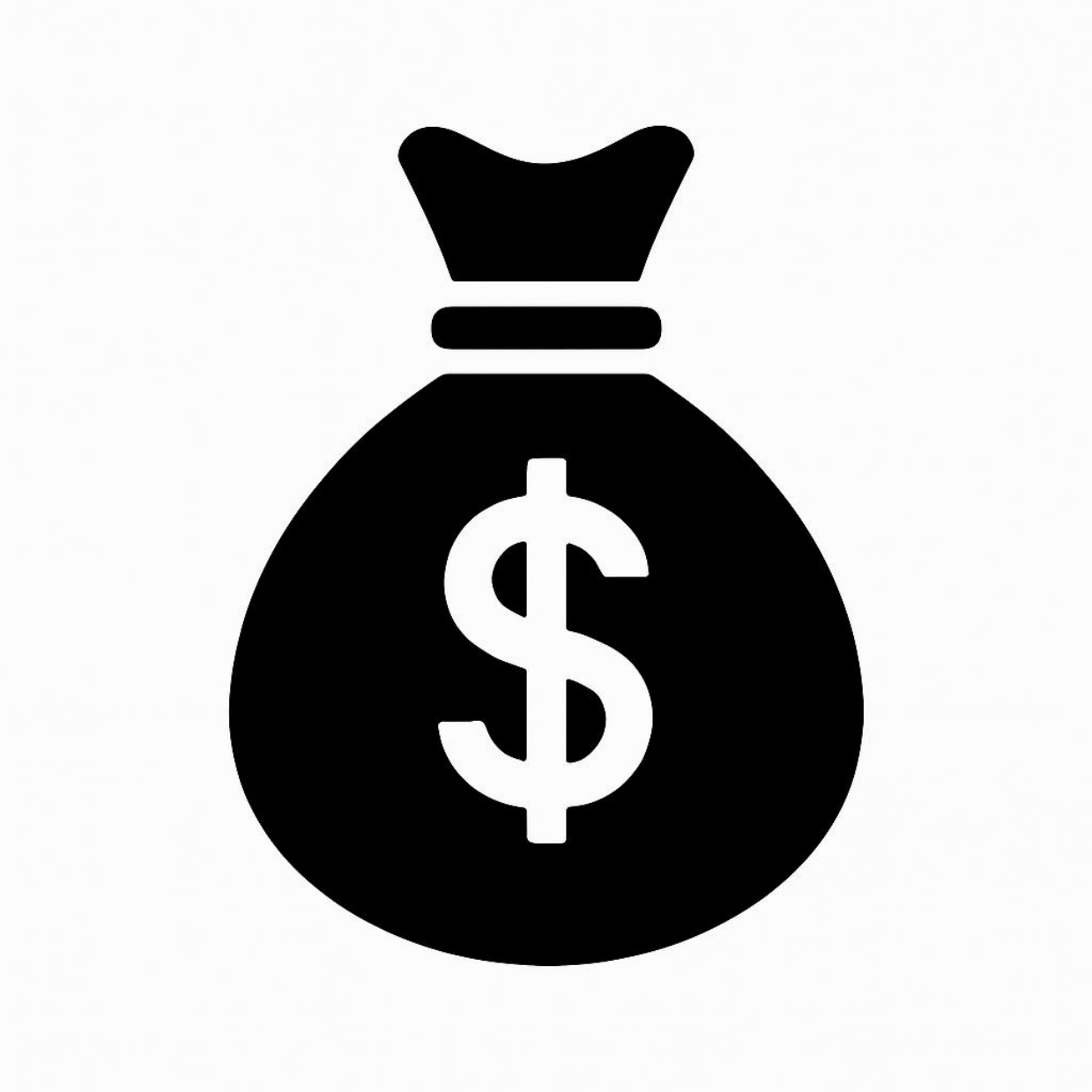} Money Bag & 6 \\
\hline
\multicolumn{3}{c}{\textit{Shapes}} \\
\hline
Shape & \ding{108} Black Circle & 13 \\
 & \ding{109} White Circle & 11 \\
 & $\square$ Square & 7 \\
 & $\triangle$ Triangle & 6 \\
 & $\lozenge$ Diamond & 6 \\
\hline
\end{tabular}
\label{tab:elementfreq}
\end{table}

\begin{table}[ht!]
\centering
\scriptsize
\caption{Top 5 intra-element co-occurrences based on unique user selections.}
\begin{tabular}{llc}
\hline
\textbf{Type} & \textbf{Co-occurring Elements} & \textbf{Frequency (Users)} \\
\hline
\multicolumn{3}{c}{\textit{Colors}} \\
\hline
Color & \colorbox[HTML]{000000}{\phantom{XX}} Black -- \colorbox[HTML]{0000FF}{\phantom{XX}} Blue & 6 \\
 & \colorbox[HTML]{000000}{\phantom{XX}} Black -- \colorbox[HTML]{FF0000}{\phantom{XX}} Red & 3 \\
 & \colorbox[HTML]{0000FF}{\phantom{XX}} Blue\hspace{0.4em} -- \colorbox[HTML]{006400}{\phantom{XX}} DarkGreen & 3 \\
 & \colorbox[HTML]{7B68EE}{\phantom{XX}} MediumSlateBlue -- \colorbox[HTML]{FF1493}{\phantom{XX}} Pink & 2 \\
 & \colorbox[HTML]{006400}{\phantom{XX}} DarkGreen --  \colorbox[HTML]{ADFF2F}{\phantom{XX}} LightGreen & 2 \\
\hline
\multicolumn{3}{c}{\textit{Icons}} \\
\hline
Icon & \includegraphics[height=1em]{img/fire-min.pdf} Fire -- \includegraphics[height=1em]{img/water_droplet2-min.pdf} Water Droplet & 4 \\
 & \includegraphics[height=1em]{img/fire-min.pdf} Fire -- \includegraphics[height=1em]{img/diamond.pdf} Gem Stone & 3 \\ 
 & \includegraphics[height=1em]{img/diamond.pdf} Gem Stone -- \includegraphics[height=1em]{img/light_bulb.pdf} Light Bulb & 3 \\
 & \includegraphics[height=1em]{img/money-bag.pdf} Money Bag -- \includegraphics[height=1em]{img/diamond.pdf} Gem Stone & 2 \\
 & \includegraphics[height=1em]{img/light_bulb.pdf} Light Bulb -- \includegraphics[height=1em]{img/money-bag.pdf} Money Bag & 2 \\ 
\hline
\multicolumn{3}{c}{\textit{Shapes}} \\
\hline
Shape & \ding{108} Black Circle -- \ding{109} White Circle & 5 \\
 & \ding{108} Black Circle -- $\square$ Square & 4 \\
 & $\square$ Square -- $\triangle$ Triangle & 3 \\
 & $\triangle$ Triangle -- $\lozenge$ Diamond & 2 \\
 & \ding{108} Black Circle -- $\lozenge$ Diamond & 2 \\
\hline
\end{tabular}
\label{tab:intraelement}
\end{table}
We investigated \emph{co-occurrences} within the same element type as well as  across different element types.
Table~\ref{tab:intraelement} summarizes the five most common intra-element co-occurrences found in participant selections.
Among colors, the combination of black and blue was the most frequently selected, chosen by six users.
This was followed by black and red, as well as blue and dark green, each selected by three users.
Less frequent pairs, such as medium slate blue and pink, and dark green and light green, were found in two user profiles each.
For icons, the pairing of fire and water droplet appeared in four instances, while combinations like fire and gem stone, and gem stone and light bulb were each observed three times.
Other icon pairs, such as money bag and gem stone, as well as light bulb and money bag, each occurred twice.
In the shape category, the pairing of black circle and white circle was the most common, appearing in five user profiles, followed by black circle and square with four user selections, and square and triangle with three.
These co-occurrence counts indicate that certain combinations of elements tend to recur across different participants.

\subsubsection{Element Arrangement}

We analyzed each element arrangement (eg., L-shape, diagonal, etc) within the 4×4 grid.
We found that at least 90\% of users placed elements within the cells indexed from 0 to 3 (top row).
In particular, 30\% place a color on index 0, 18\% place an icon at index 1, and another 18\% place a shape at index 2.
We investigated the arrangement pattern for each user and, when multiple patterns existed, considered only the longest one.
Table~\ref{tab:pattern_summary} presents the results.

\begin{table}[ht]
\centering
\small
\caption{Frequent Patterns.}
\label{tab:pattern_summary}
\begin{tabular}{lcccc}
\hline
\multirow{2}{*}{\textbf{Pattern Type}} & \multicolumn{4}{c}{\textbf{Frequency}} \\ \cline{2-5}
 & \textbf{$<$20} & \textbf{20--30} & \textbf{$>$30} & \textbf{Total} \\
\hline
Horizontal Line & 7 & 11 & 5 & 23 \\
L-shape         & 1 & 13 & 5 & 19 \\
Diagonal        & 1 & 7  & 1 & 9 \\
Square 2$\times$2 & 0 & 3  & 1 & 4 \\
Square 3$\times$3 & 0 & 1  & 1 & 2 \\
Undefined       & 1 & 1  & 0 & 2 \\
\hline
\textbf{Total}  & 10 & 36 & 13 & 59 \\
\hline
\end{tabular}
\end{table}

The most frequent patterns were horizontal and L-shaped, with total frequencies of 23 and 19, respectively.
Square-based patterns appeared far less frequently, with only four instances of 2×2 squares and two of 3×3 squares.

When analyzed by age group, participants under 20 years old predominantly formed horizontal lines, with 7 out of 10 patterns following this configuration. Participants aged 20–30 exhibited the greatest diversity in pattern types, suggesting a more explorative behavior. Within this group, L-shaped configurations were the most frequent (13 occurrences), followed by horizontal lines (11) and diagonals (7). Patterns observed in the over-30 group resembled those of the under-20 group, with horizontal and L-shaped patterns being the most common (5 occurrences each).
These results indicate that most participants preferred easily reproducible patterns. These results indicate that most participants preferred easily reproducible patterns. 

For each secret, we recorded the grid cells a participant selected and the elements they placed in those cells. Every distinct combination of these attributes counts as a unique secret. We then grouped matching secrets across participants and counted how often each one appeared.
From these counts, we computed probabilities
$
p_i = \frac{c_i}{N},
$
where \(c_i\) is the count of secret \(i\) and \(N\) is the total number of secrets. Using these probabilities, we calculated the empirical Shannon entropy
$
H = -\sum_i p_i \log_2 p_i .
$
In total, we identified 59 secrets. Fifty-seven appeared only once, and one secret occurred twice. The repeated secret therefore has probability \(2/59\), while each remaining secret has probability \(1/59\). Substituting these values into the entropy formula yields an empirical entropy of 5.85 bits. This result reaches about $99.4\%$ of the sample-based maximum, as the empirical entropy estimate cannot exceed $\log_2(59)\approx 5.88$ bits.
We treat the empirical entropy as a sample-based diversity indicator (bounded by $\log_2 N$), not as evidence of strong real-world resistance, particularly given the observed preference for simple pattern families that may concentrate choices at larger scale.
%

To describe resistance to offline guessing, we also report guessability. Guessability captures how many guesses an attacker needs when they try the most common user-chosen secrets first~\cite{malone2005}. We take each participant's registered secret, count how often each distinct secret occurs, and then sort the secrets from most to least frequent. Let $p_i$ denote the observed probability of the $i$-th most frequent secret. Our data contains $m=58$ unique secrets. One secret occurs twice, so $p_1=2/59$. The remaining $57$ secrets occur once, so $p_i=1/59$ for $i=2,\dots,58$. We report the expected guesswork, which measures the average number of guesses needed to recover a randomly chosen user secret under this frequency-based strategy:
\[
E[G] = \sum_{i=1}^{58} i \cdot p_i
      = 1\cdot\frac{2}{59} + \sum_{i=2}^{58} i\cdot\frac{1}{59}
      = \frac{1712}{59} \approx 29.02
\]
We also report an $\alpha$-work factor, which measures how many top-ranked guesses allow an attacker to compromise an $\alpha$ fraction of accounts:
\[
W(\alpha) = \min\left\{k : \sum_{i=1}^{k} p_i \ge \alpha \right\}
\]
With $p_1=2/59$ and $p_i=1/59$ for $i\ge 2$, the cumulative coverage is $S_1=2/59$ and $S_k=(k+1)/59$ for $k\ge 2$. This yields $W(0.10)=5$, $W(0.20)=11$, and $W(0.50)=29$. In practical terms, an attacker who guesses secrets in descending frequency needs 5 guesses to reach at least 10\% success, 11 guesses to reach at least 20\% success, and 29 guesses to reach at least 50\% success on this participant set.

\subsection{Lessons Learned}

The voluntary nature of this experiment, along with the lack of consequences for using simple secrets, may have influenced the results. Nonetheless, this preliminary study yielded several valuable insights.

Participants often have preferences for specific elements, which can limit the variety of secret selections. To address this bias, it is beneficial to expand the set of available elements and present a randomized selection during registration, while preserving per-service consistency so users can reproduce the same element set at login.
Nonetheless, managing distinct secrets across multiple services remains a practical usability tension.

We observe that participants choose reproducible patterns.
Future iterations could include lightweight registration-time guidance to discourage common pattern types (\eg straight lines) and encourage broader spatial coverage. In addition, the interface could present a simple ``diversity'' indicator based on local heuristics (\eg number of rows/columns used or direction changes), which may help steer users toward less concentrated choices while maintaining memorability.

The visual elements used in the scheme should be selected with care. For example, color-based elements can pose challenges for users with color vision deficiencies, an issue we had not fully recognized before this study. A wide range of alternative visual elements can better accommodate diverse user groups, including localized alphabets, cartoon characters, textured symbols, profession-specific icons, and other easily distinguishable sets.
We anticipate a strong potential for offering this authentication scheme to specific user groups, such as children, who may perceive it as a game while gaining protection against weak passwords~\cite{assal2018}. 

%

Finally, we need a ``forgot password'' feature to measure two metrics: the False Rejection Rate (FRR), which shows how often the system denies access to a legitimate user, and the False Acceptance Rate (FAR), which shows how often the system grants access to an impostor.

\section{Threat to Validity}
\label{sec:ttv}
\emph{Threat to Internal Validity} refers to the variation in user environments, as participants completed the registration and login phases remotely using their own devices.
These differences may influence how users interact with the interface, especially regarding screen size, input methods, and potential distractions during the tasks.
%
In addition, we did not collect reasons for drop-out and did not collect post-study satisfaction, which will be investigated 
in future studies.

\emph{Threat to External Validity} refers to the generalizability of the study's findings to broader populations and different contexts.
While participants from various age groups were included, most were recruited from the researchers’ networks and communities, which may not accurately reflect a wider or more diverse user base.

\section{Conclusion}
\label{sec:conclusion}

We presented a graphical authentication scheme called \emph{Pick and Sort}.
Each login attempt requires only a few simple clicks or touches, where users choose a set of visual elements and arrange them within a grid. The setup can be customized by selecting different elements and adjusting the grid size to better accommodate various user groups.

Initial results show that the scheme is slower to complete than conventional methods. However, it may be appropriate where login speed is not the primary requirement, such as infrequent-access use cases or as a secondary authentication mechanism. 

Future work includes examining user behavior in real-world settings, studying specific age groups such as children, considering long-term memorability, measuring performance across devices and environments, exploring the resistance of the scheme to guessing attacks, and improving the scheme’s scalability across multiple services.

\bibliographystyle{ACM-Reference-Format}
\bibliography{main}

\end{document}